# A New Advanced User Authentication and Confidentiality Security Service

Sanjay Majumder
Institute Of Engineering & Management
(IEM) Salt Lake Sec V,
Kolkata, West Bengal, India

Sanjay Chakraborty
Professor, Institute Of Engineering & Management
(IEM) Salt Lake Sec V,
Kolkata, West Bengal, India

Suman Das
Institute Of Engineering & Management
(IEM) Salt Lake Sec V,
Kolkata, West Bengal, India

## ABSTRACT
Network & internet security is the burning question of today's world and they are deeply related to each other for secure successful data transmission. Network security approach is totally based on the concept of network security services. In this paper, a new system of network security service is implemented which is more secure than conventional network security services. This technique is mainly deals with two essential network security services, one is user authentication and other is data confidentiality. For user authentication this paper introduces 'Graphical Username' & 'Voice Password' approaches which provides better security than conventional 'username '& 'password' authentication process. In data confidentiality section this paper introduces two layer private key for both message encryption & decryption which is mainly applicable on 8 bit plain text data. This paper also provides the hints of introducing other two network security services (integrity and non-repudiation) as a future work.

## 1. INTRODUCTION
Computing & internet has become the integral part of our day to day life. Everyday various information are exchanged over the internet. So security of the information is highly sensitive issue now-a-days. For the security of internet & network security there are four basic principles. They are Authentication, Confidentiality, Integrity & Non-repudiation. User authentication is referred as 'Provision of Assurance that the message is originated from Authorized user' [2]. Data confidentiality refers to limiting information access & disclosure to authorized user and preventing access or disclosure to unauthorized one. For data confidentiality cryptography is used to a great extent. Cryptography is the art of achieving security& hiding information from unauthorized person. Cryptography is the combination of mathematics & computer science. There are two types of cryptography – Public key cryptography & private key cryptography [5]. Integrity refers the sent message is not tempered & altered. Non-repudiation is the assurance that someone cannot deny something. Typically, non-repudiation refers to the ability to ensure that a party to a contact or a communication cannot deny the authenticity of their signature on a document or the sending of a message that they originated. Conventional cryptography & network security system are not handy enough to secure the network. Conventional user name & password for user authentication is not totally secure. Short length of password has the risk of hacking & lengthy alpha - numeric password is difficult to memorize. On the other hand, one private key used for data encryption & decryption is not enough secure.

The rest of this paper is organized as follows. Section 2 discusses related works on these both user authentication & data confidentiality techniques. The proposed technology is discussed is discussed in section 3. Section 4 describes user authentication. Subsection4.1 of section 4 describes authentication through graphical user name. Subsection4.2 of section 4 describes authentication through voice password. Section 5 describes data confidentiality. Subsection5.1 of section 5 describes key generation. Subsection5.2 of section 5 describes plain text to cipher text conversion. Subsection5.3 of section 5 describes cipher text to plain text conversion. Section 6 concludes with a summary of the paper, area of implication & future work. 'References' finally follows the conclusion.

## 2. RELATED WORK
Lot of works has been done on user authentication & data confidentiality. They are also famous for their easy to use nature.

A paper describes method & techniques of user authentication. This paper analyze alpha-numeric password, biometric password, token password & centralized authentication system [2].

A paper analyzes the factors & issues of voice recognition [4].

A paper investigates the feasibility study on graphics card for cryptography. This paper uses symmetric key along with computer graphics card for data confidentiality [6].

A paper based on private key cryptography which uses ASCII values to convert the plain text to cipher text [5].

## 3. PROPOSED METHODOLOGY
This paper introduces a new & unique way of user authentication & data confidentiality. User authentication is done through two factor authentication system. First factor is graphical user name. Second factor is voice password. Data confidentiality is done through public key cryptography. Two keys are used in different phase of plain text to cipher text & cipher text to plain text conversion process. Two layer keys make the data more secure & confidential.

User Authentication:

Select 3 pictures

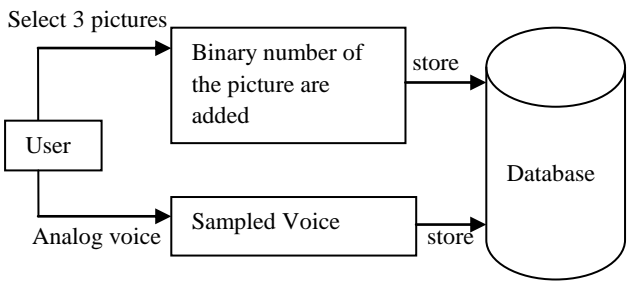

**Fig.1: User Authentication.**

Data Confidentiality (After user authentication):

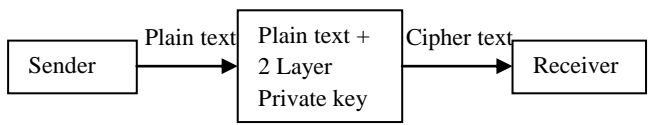

**Fig.2: Data Confidentiality.**

## 4. USER AUTHENTICATION

User authentication is done through two processes. First process is graphical user name & second process is voice password.

### 4.1 User authentication through graphical user name

User signup algorithm:

1. System will provide a series of pictures i.e. 50 pictures.

2. 3 pictures are randomly selected by the user & user has to remember the sequence. These 3 pictures will act like a user name.

3. Every picture has 8 bit unique binary number i.e. 10001011, 01001010, 11010001.

4. System will accumulate the binary numbers i.e. 100010110100101011010001 & save it as a pattern.

5. System will save the pattern in the database for future login

User login Algorithm:

1. To login the user has to select previously chosen pictures sequentially.

2. The system will match the pattern with previously saved pattern.

3. If the current pattern is matched with the previously saved pattern, system will allow the user to log in.

4. If the current pattern does not match with the saved pattern, the user is refused to login by the system.

User authentication through graphical user name flowchart:

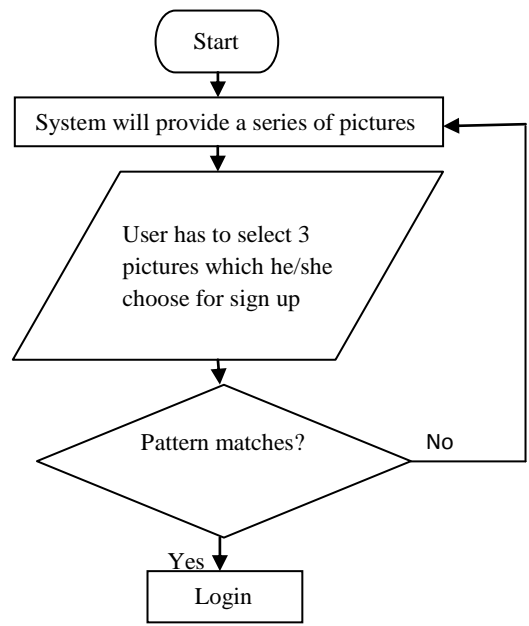

**Fig.3: Flow chart of user authentication trough user name.**

### 4.2 User authentication through voice password

User signup algorithm:

1. System will provide a button to record the voice.

2. User has to click on the button & say something.

3. System will record the voice & sample the sound in digital format (binary form).

4. System will save the binary number in the database which will act as password.

User signup flowchart:

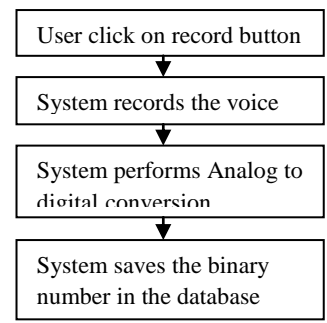

**Fig. 4: Flow chart user authentication through voice password (signup).**

User login algorithm:

1. System will provide a button to record the voice.

2. User has to click on the button & say the previously recorded word.

3. System will capture the sound.

4. System will sample the sound & digitized in binary form.

5. Then system will match both the binary pattern.

6. If the current pattern is matched with the previously saved pattern, system will allow the user to log in.
7. If the current pattern does not match with the saved pattern, the user is refused to login by the system.

User login flowchart:

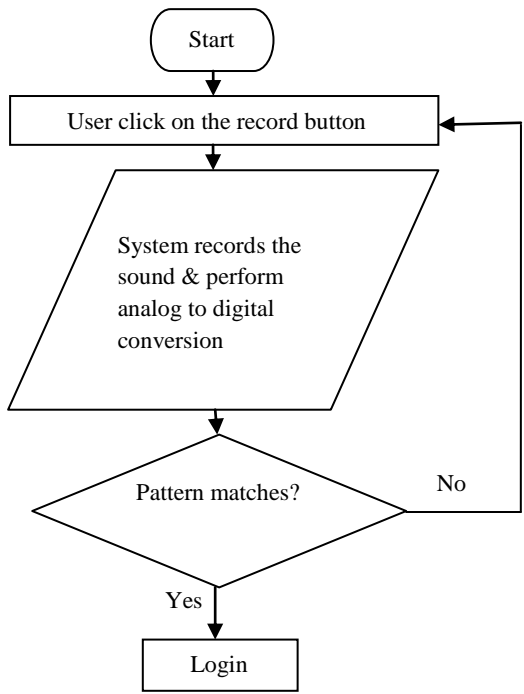

**Fig. 5: Flow chart user authentication through voice password (login).**

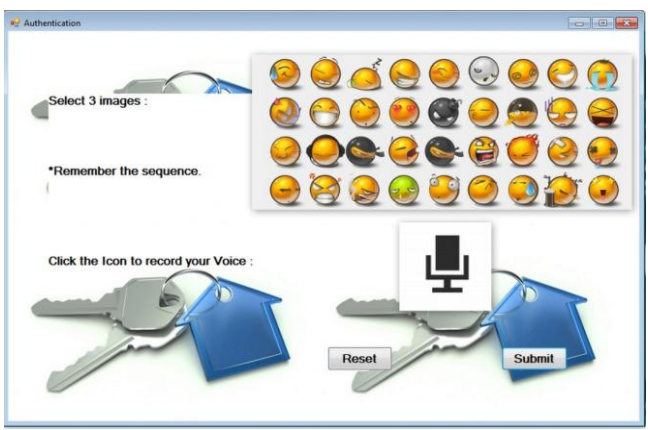

**Fig 6: Snapshot of authentication system.**

## 5. CONFIDENTIALITY
Two keys are used for both encryption & decryption purpose.

## 5.1 Key Generation
*5.1.1 Key 1 generation*
Key 1 is generated through the following steps:

1. A 10 bit number is entered & stored in an array.
2. A permutation operation is performed on the array.
   - $3^{rd}$ bit of the array comes in $1^{st}$ bit position.
   - $5^{th}$ bit of the array comes in $2^{nd}$ bit position.
   - $2^{nd}$ bit of the array comes in $3^{rd}$ bit position.
   - $7^{th}$ bit of the array comes in $4^{th}$ bit position.
   - $4^{th}$ bit of the array comes in $5^{th}$ bit position.
   - $10^{th}$ bit of the array comes in $6^{th}$ bit position.
   - $1^{st}$ bit of the array comes in $7^{th}$ bit position.
   - $9^{th}$ bit of the array comes in $8^{th}$ bit position.
   - $8^{th}$ bit of the array comes in $9^{th}$ bit position.
   - $6^{th}$ bit of the array comes in $10^{th}$ bit position.
3. The number after permutation is stored in an array.
4. The first 5 bits of the array is taken in a different array & last 5 bits is taken in another array.
5. Left shift operation is performed on both of the array.
6. After the left shift operation the two arrays are merged & one single array is formed out of the two arrays.
7. This new array undergoes a permutation operation
   - $6^{th}$ bit of the array comes in $1^{st}$ bit position.
   - $3^{rd}$ bit of the array comes in $2^{nd}$ bit position.
   - $7^{th}$ bit of the array comes in $3^{rd}$ bit position.
   - $4^{th}$ bit of the array comes in $4^{th}$ bit position.
   - $8^{th}$ bit of the array comes in $5^{th}$ bit position.
   - $5^{th}$ bit of the array comes in $6^{th}$ bit position.
   - $10^{th}$ bit of the array comes in $7^{th}$ bit position.
   - $9^{th}$ bit of the array comes in $8^{th}$ bit position.
8. The number after permutation is stored in an array & this number is treated as key1.

*5.1.2 Key 2 generation*
Key 2 is generated through the following steps:

1. A 10 bit number is entered & stored in an array.
2. A permutation operation is performed on the array.
   - $3^{rd}$ bit of the array comes in $1^{st}$ bit position.
   - $5^{th}$ bit of the array comes in $2^{nd}$ bit position.
   - $2^{nd}$ bit of the array comes in $3^{rd}$ bit position.
   - $7^{th}$ bit of the array comes in $4^{th}$ bit position.
   - $4^{th}$ bit of the array comes in $5^{th}$ bit position.
   - $10^{th}$ bit of the array comes in $6^{th}$ bit position.
   - $1^{st}$ bit of the array comes in $7^{th}$ bit position.
   - $9^{th}$ bit of the array comes in $8^{th}$ bit position.
   - $8^{th}$ bit of the array comes in $9^{th}$ bit position.
   - $6^{th}$ bit of the array comes in $10^{th}$ bit position.
3. The number after permutation is stored in an array.
4. The first 5 bits of the array is taken in a different array & last 5 bits is taken in another array.
5. Left shift operation is performed on both of the array.
6. After the left shift operation the two arrays are merged & one single array is formed out of the two arrays.
7. This new array undergoes a permutation operation
   - $6^{th}$ bit of the array comes in $1^{st}$ bit position.

- ➢ 3$^{rd}$ bit of the array comes in 2$^{nd}$ bit position.
- ➢ 7$^{th}$ bit of the array comes in 3$^{rd}$ bit position.
- ➢ 4$^{th}$ bit of the array comes in 4$^{th}$ bit position.
- ➢ 8$^{th}$ bit of the array comes in 5$^{th}$ bit position.
- ➢ 5$^{th}$ bit of the array comes in 6$^{th}$ bit position.
- ➢ 10$^{th}$ bit of the array comes in 7$^{th}$ bit position.
- ➢ 9$^{th}$ bit of the array comes in 8$^{th}$ bit position.

8. The number after permutation is stored in an array & this number is treated as key2.

Flow chart of key generation:

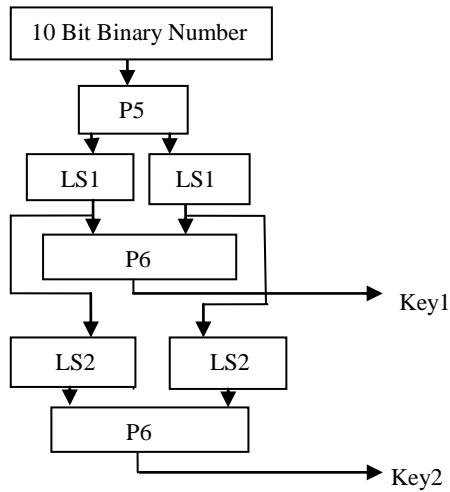

**Fig. 7: Flow chat of key generation.**

## 5.2 Conversion of Plain Text to Cipher Text

The conversion procedure of plain text to cipher text occurs through the following steps:

1. The user enters 8 bit plain text as input which will be stored in an array array1.

2. A permutation operation is performed on that 8 bit text
   - ➢ 2$^{nd}$ bit of the plain text comes in 1$^{st}$ position.
   - ➢ 6$^{th}$ bit of the plain text comes in 2$^{nd}$ position.
   - ➢ 3$^{rd}$ bit of the plain text comes in 3$^{rd}$ position.
   - ➢ 1$^{st}$ bit of the plain text comes in 4$^{th}$ position.
   - ➢ 5$^{th}$ bit of the plain text comes in 5$^{th}$ position.
   - ➢ 8$^{th}$ bit of the plain text comes in 6$^{th}$ position.
   - ➢ 4$^{th}$ bit of the plain text comes in 7$^{th}$ position.
   - ➢ 7$^{th}$ bit of the plain text comes in 8$^{th}$ position.

3. The text after permutation is stored in an array2.

4. The first 4 bits of the newly formed array array2 is taken in a different array3 & last 4bits of the array2 is taken in another array array4.

5. A permutation is performed on the array4 containing the last 4 bits of array2
   - ➢ 4$^{th}$ bit of the array comes in 1$^{st}$ position.
   - ➢ 1$^{st}$ bit of the array comes in 2$^{nd}$ position.
   - ➢ 2$^{nd}$ bit of the array comes in 3$^{rd}$ position.
   - ➢ 3$^{rd}$ bit of the array comes in 4$^{th}$ position.
   - ➢ 2$^{nd}$ bit of the array comes in 5$^{th}$ position.
   - ➢ 3$^{rd}$ bit of the array comes in 6$^{th}$ position.
   - ➢ 4$^{th}$ bit of the array comes in 7$^{th}$ position.
   - ➢ 1$^{st}$ bit of the array comes in 8$^{th}$ position.

6. The number after permutation is stored in an array array5

7. An 8 bit key key1 is generated.

8. An XOR operation is performed between the array5 & key1, the result is stored in an array array6.

9. The 1$^{st}$ & 4$^{th}$ bit of array6 is stored in 1$^{st}$ & 2$^{nd}$ bit position of a different array array7 and 2$^{nd}$ & 3$^{rd}$ bit of the array6 is stored in 1$^{st}$ & 2$^{nd}$ bit position of another array array8

10. The binary number in array7 is converted to the corresponding decimal number & the result is stored into a variable (bin1) & the binary number in array8 is converted to the corresponding decimal number & the result is stored into a variable (bin2).

11. From the matrix S1, the value corresponding to S1 [bin1][bin2] is taken & stored in a variable(var1).

12. The decimal number in var1 is converted to 2 bit binary number & stored in an array array9.

13. The 5$^{th}$ & 8$^{th}$ bit of array6 is stored in 1$^{st}$ & 2$^{nd}$ bit position of array array10 & 6$^{th}$ & 7$^{th}$ bit of array6 is stored in 1$^{st}$ & 2$^{nd}$ bit position of an array array11.

14. The binary number in array10 is converted to the corresponding decimal number & the result is stored into a variable (bin3) & the binary number in array11 is converted to the corresponding decimal number & the result is stored into a variable (bin4).

15. From the matrix S2, the value corresponding to S2 [bin3][bin4] is taken & stored in a variable (var2).

16. The decimal number in var2 is converted to 2 bit binary number & stored in an array array12.

17. The two array array9 & array12 is merged to form a new array array13 of 4bit.

18. A permutation is performed on the array13
    - ➢ 2$^{nd}$ bit of the array comes in 1$^{st}$ bit position.
    - ➢ 4$^{th}$ bit of the array comes in 2$^{nd}$ bit position.
    - ➢ 3$^{rd}$ bit of the array comes in 3$^{rd}$ bit position.
    - ➢ 1$^{st}$ bit of the array comes in 4$^{th}$ bit position.

19. The result after permutation is stored in an array array14.

20. An XOR operation is performed between array14 & array3 & the 4 bit result is stored in an array array15

21. A permutation is performed on the array15
    - ➢ 4$^{th}$ bit of the array comes in 1$^{st}$ position.
    - ➢ 1$^{st}$ bit of the array comes in 2$^{nd}$ position.
    - ➢ 2$^{nd}$ bit of the array comes in 3$^{rd}$ position.
    - ➢ 3$^{rd}$ bit of the array comes in 4$^{th}$ position.
    - ➢ 2$^{nd}$ bit of the array comes in 5$^{th}$ position.
    - ➢ 3$^{rd}$ bit of the array comes in 6$^{th}$ position.
    - ➢ 4$^{th}$ bit of the array comes in 7$^{th}$ position.
    - ➢ 1$^{st}$ bit of the array comes in 8$^{th}$ position.

22. The result after permutation is stored in an array array16.

23. A new key key2 is generated.

24. An XOR operation is performed between key2 & array16 & the result is stored in an array array17.

25. The $1^{st}$ & $4^{th}$ bit of array17 is stored in $1^{st}$ & $2^{nd}$ bit position of array array18 & $2^{nd}$ & $3^{rd}$ bit of array17 is stored in $1^{st}$ & $2^{nd}$ bit position of an array array19.

26. The binary number in array18 is converted to the corresponding decimal number & the result is stored into a variable (bin5) & the binary number in array19 is converted to the corresponding decimal number & the result is stored into a variable (bin6).

27. From the matrix S1, the value corresponding to S1 [bin5][bin6] is taken & stored in a variable (var3).

28. The decimal number in var3 is converted to 2 bit binary number & stored in an array array20.

29. The $5^{th}$ & $8^{th}$ bit of array17 is stored in $1^{st}$ & $2^{nd}$ bit position of array array21 & $6^{th}$ & $7^{th}$ bit of array17 is stored in $1^{st}$ & $2^{nd}$ bit position of an array array22.

30. The binary number in array21 is converted to the corresponding decimal number & the result is stored into a variable (bin7) & the binary number in array22 is converted to the corresponding decimal number & the result is stored into a variable (bin8).

31. From the matrix S2, the value corresponding to S2 [bin7][bin8] is taken & stored in a variable (var4).

32. The decimal number in var4 is converted to 2 bit binary number & stored in an array array23.

33. The two array array20 & array23 are merged & form a new array array24 of 4 bit.

34. A permutation is performed on the array24.

    ➢ $2^{nd}$ bit of the array comes in $1^{st}$ bit position.
    ➢ $4^{th}$ bit of the array comes in $2^{nd}$ bit position.
    ➢ $3^{rd}$ bit of the array comes in $3^{rd}$ bit position.
    ➢ $1^{st}$ bit of the array comes in $4^{th}$ bit position.

35. The number after permutation is stored in an array array25.

36. An XOR operation is performed between array25 & array4 & the result is stored in an array array26.

37. The two array array16 & array15 are merged & a new array array27 is formed.

38. A permutation operation is formed on array27

    ➢ $4^{th}$ bit of the array comes in $1^{st}$ bit position.
    ➢ $1^{st}$ bit of the array comes in $2^{nd}$ bit position.
    ➢ $3^{rd}$ bit of the array comes in $3^{rd}$ bit position.
    ➢ $5^{th}$ bit of the array comes in $4^{th}$ bit position.
    ➢ $7^{th}$ bit of the array comes in $5^{th}$ bit position.
    ➢ $2^{nd}$ bit of the array comes in $6^{th}$ bit position.
    ➢ $8^{th}$ bit of the array comes in $7^{th}$ bit position.
    ➢ $6^{th}$ bit of the array comes in $8^{th}$ bit position.

39. The result is stored in an array array28 & array28 is the cipher text of array1 containing the plain text.

Flow chart of Plain text to cipher text conversion:

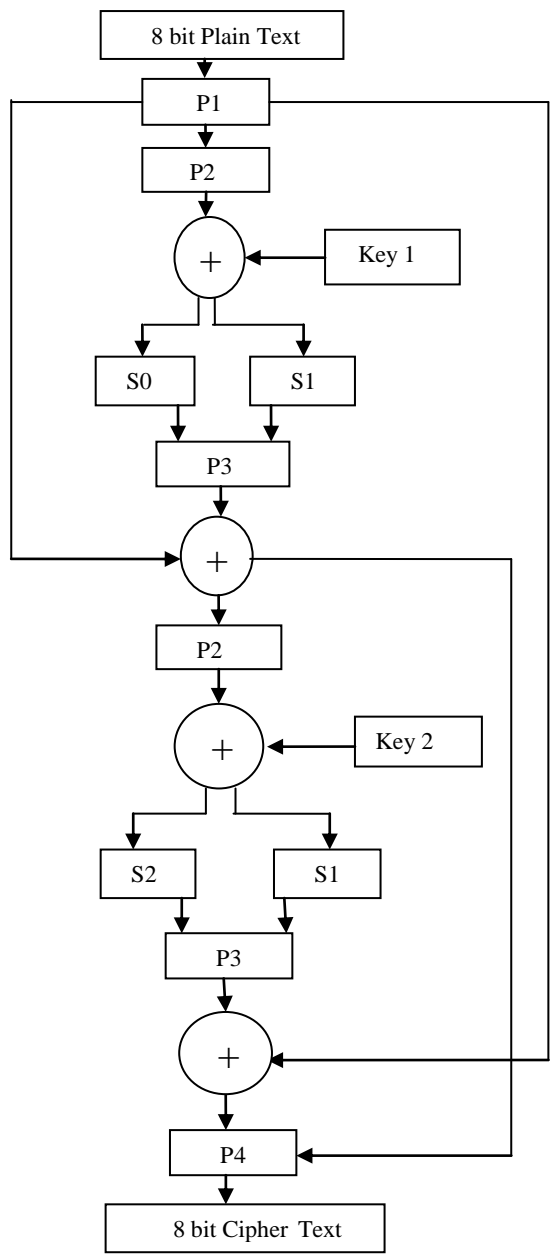

**Fig. 8: Flow chat of plain text to cipher text conversion.**

## 5.3 Conversion of Cipher Text to plain Text
The conversion procedure of plain text to cipher text occurs through the following steps:

1. The user enters 8 bit cipher text as input which will be stored in an array array1.

2. A permutation operation is performed on that 8 bit text.

    ➢ $2^{nd}$ bit of the cipher text comes in $1^{st}$ position.
    ➢ $6^{th}$ bit of the cipher text comes in $2^{nd}$ position.
    ➢ $3^{rd}$ bit of the cipher text comes in $3^{rd}$ position.
    ➢ $1^{st}$ bit of the cipher text comes in $4^{th}$ position.
    ➢ $5^{th}$ bit of the cipher text comes in $5^{th}$ position.
    ➢ $8^{th}$ bit of the cipher text comes in $6^{th}$ position.

- ➢ 4th bit of the cipher text comes in 7th position.
- ➢ 7th bit of the cipher text comes in 8th position.

3. The text after permutation is stored in an array2.
4. The first 4 bits of the newly formed array array2 is taken in a different array3 & last 4bits of the array2 is taken in another array array4.
5. A permutation is performed on the array4 containing the last 4 bits of array2
   - ➢ 4th bit of the array comes in 1st position.
   - ➢ 1st bit of the array comes in 2nd position.
   - ➢ 2nd bit of the array comes in 3rd position.
   - ➢ 3rd bit of the array comes in 4th position.
   - ➢ 2nd bit of the array comes in 5th position.
   - ➢ 3rd bit of the array comes in 6th position.
   - ➢ 4th bit of the array comes in 7th position.
   - ➢ 1st bit of the array comes in 8th position.
6. The number after permutation is stored in an array array5
7. An 8 bit key key1 is generated.
8. An XOR operation is performed between the array5 & key1, the result is stored in an array array6.
9. The 1st & 4th bit of array6 is stored in 1st & 2nd bit position of a different array array7 and 2nd & 3rd bit of the array6 is stored in 1st & 2nd bit position of another array array8
10. The binary number in array7 is converted to the corresponding decimal number & the result is stored into a variable (bin1) & the binary number in array8 is converted to the corresponding decimal number & the result is stored into variable (bin2).
11. From the matrix S1, the value corresponding to S1 [bin1][bin2] is taken & stored in a variable(var1).
12. The decimal number in var1 is converted to 2 bit binary number & stored in an array array9.
13. The 5th & 8th bit of array6 is stored in 1st & 2nd bit position of array array10 & 6th & 7th bit of array6 is stored in 1st & 2nd bit position of an array array11.
14. The binary number in array10 is converted to the corresponding decimal number & the result is stored into a variable (bin3) & the binary number in array11 is converted to the corresponding decimal number & the result is stored into a variable (bin4).
15. From the matrix S2, the value corresponding to S2 [bin3][bin4] is taken & stored in a variable (var2).
16. The decimal number in var2 is converted to 2 bit binary number & stored in an array array12.
17. The two array array9 & array12 is merged to form a new array array13 of 4bit.
18. A permutation is performed on the array13
    - ➢ 2nd bit of the array comes in 1st bit position.
    - ➢ 4th bit of the array comes in 2nd bit position.
    - ➢ 3rd bit of the array comes in 3rd bit position.
    - ➢ 1st bit of the array comes in 4th bit position.

19. The result after permutation is stored in an array array14.
20. An XOR operation is performed between array14 & array3 & the 4 bit result is stored in an array array15
21. A permutation is performed on the array15
    - ➢ 4th bit of the array comes in 1st position.
    - ➢ 1st bit of the array comes in 2nd position.
    - ➢ 2nd bit of the array comes in 3rd position.
    - ➢ 3rd bit of the array comes in 4th position.
    - ➢ 2nd bit of the array comes in 5th position.
    - ➢ 3rd bit of the array comes in 6th position.
    - ➢ 4th bit of the array comes in 7th position.
    - ➢ 1st bit of the array comes in 8th position.
22. The result after permutation is stored in an array array16.
23. A new key key2 is generated.
24. An XOR operation is performed between key2 & array16 & the result is stored in an array array17.
25. The 1st & 4th bit of array17 is stored in 1st & 2nd bit position of array array18 & 2nd & 3rd bit of array17 is stored in 1st & 2nd bit position of an array array19.
26. The binary number in array18 is converted to the corresponding decimal number & the result is stored into a variable (bin5) & the binary number in array19 is converted to the corresponding decimal number & the result is stored into a variable (bin6).
27. From the matrix S1, the value corresponding to S1 [bin5][bin6] is taken & stored in a variable (var3).
28. The decimal number in var3 is converted to 2 bit binary number & stored in an array array20.
29. The 5th & 8th bit of array17 is stored in 1st & 2nd bit position of array array21 & 6th & 7th bit of array17 is stored in 1st & 2nd bit position of an array array22.
30. The binary number in array21 is converted to the corresponding decimal number & the result is stored into a variable (bin7) & the binary number in array22 is converted to the corresponding decimal number & the result is stored into a variable (bin8).
31. From the matrix S2, the value corresponding to S2 [bin7][bin8] is taken & stored in a variable (var4).
32. The decimal number in var4 is converted to 2 bit binary number & stored in an array array23.
33. The two array array20 & array23 are merged & form a new array array24of 4 bit.
34. A permutation is performed on the array24.
    - ➢ 2nd bit of the array comes in 1st bit position.
    - ➢ 4th bit of the array comes in 2nd bit position.
    - ➢ 3rd bit of the array comes in 3rd bit position.
    - ➢ 1st bit of the array comes in 4th bit position.
35. The number after permutation is stored in an array array25.
36. An XOR operation is performed between array25 & array4 & the result is stored in an array array26.

37. The two array array16 & array15 are merged & a new array array27 is formed.
38. A permutation operation is formed on array27
    - 4th bit of the array comes in 1st bit position.
    - 1st bit of the array comes in 2nd bit position.
    - 3rd bit of the array comes in 3rd bit position.
    - 5th bit of the array comes in 4th bit position.
    - 7th bit of the array comes in 5th bit position.
    - 2nd bit of the array comes in 6th bit position.
    - 8th bit of the array comes in 7th bit position.
    - 6th bit of the array comes in 8th bit position.
39. The result is stored in an array array28 & array28 is the plain text of array1 containing the cipher text.

Flow Chart of cipher text to plain text conversion:

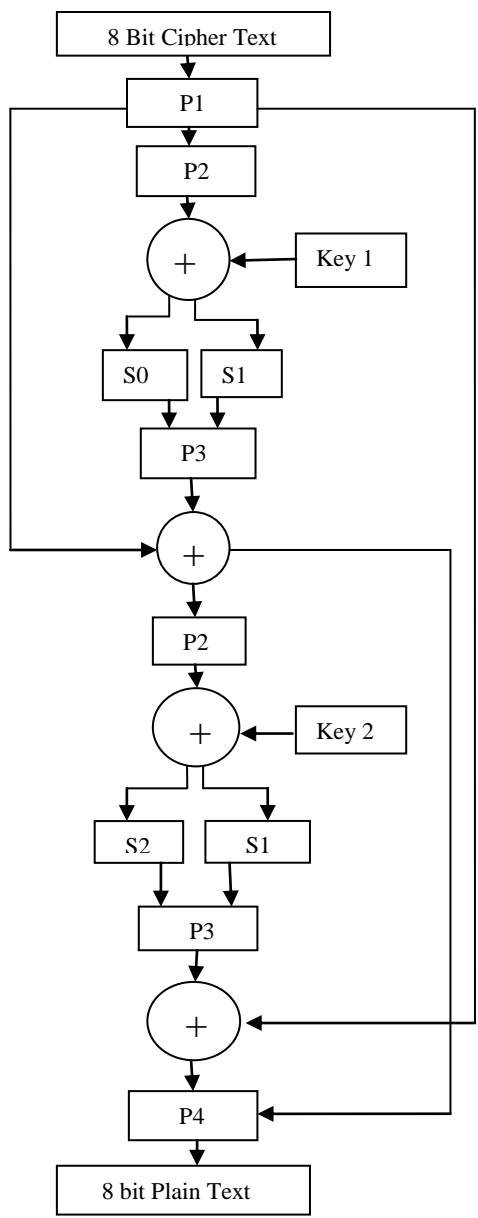

**Fig. 9: Flow chat of cipher text to plain text conversion.**

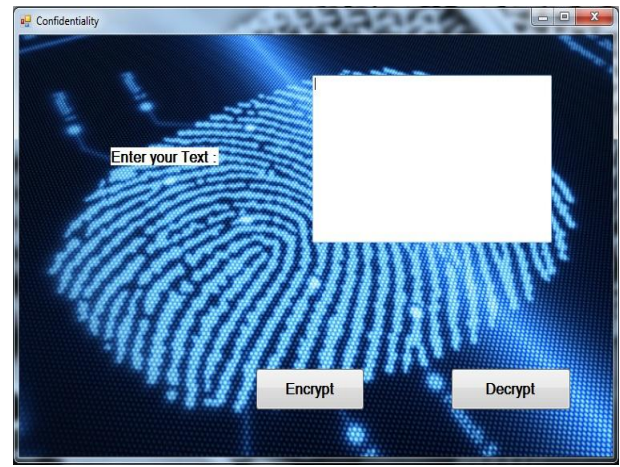

**Fig 10: Snapshot of data confidentiality system.**

## 6. PERFORMANCE EVALUTION
In this paper, multimedia objects are used for authentication which cause more time & space complexity than tradition user name & password system. In data confidentiality, two keys are used in two different steps in both encryption & decryption which also cause more time complexity than traditional system. But in the point of security parameter, this system is more secure than traditional system.

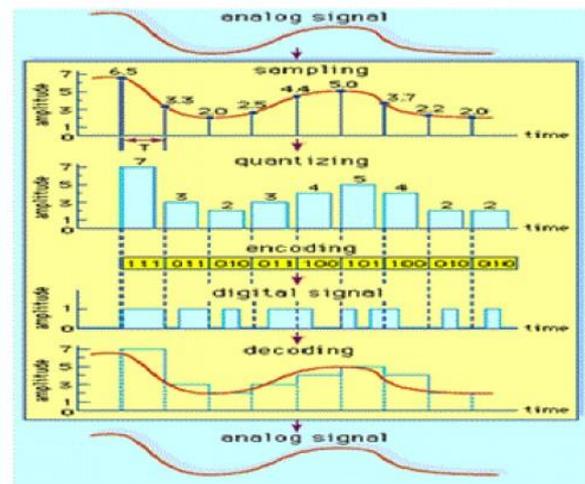

**Fig 11: An example of analog to digital conversion**

## 7. CONCLUSION & FUTURE WORK
Network security is a challenging task for data transmission now a day. It actually motivates the researcher to propose more secure techniques to prevent those challenges during data transmission. Motivating from this issue, the main research aim of this paper is introduced. It proposes a new user authentication and data confidentiality service which results much better service in terms of security than the conventional services.

In future, this concept may be extended to introduce the other services such as integrity and non-repudiation to build a complete network security software service which helps to provide an entire security for data transmission. This approach can be also applied on cloud architecture to provide secure cloud data transmission.

## 8. ACKNOWLEDGMENTS
Special thanks to Prof. Sourav Saha, Prof. Ee-Kian Wong, Prof. Pinaki Karmakar and Mr. Uttaran Bhattacharya whose advice & technical help improved the presentation of this article.

## 10. AUTHOR'S PROFILE
**Sanjay Majumder:** B. Tech. from West Bengal University of Technology, India on Computer Science & Engineering in the year 2014. His areas of interests are Music Technology and Network Security.

**Sanjay Chakraborty:** B-Tech from West Bengal University of Technology, India on Information Technology in the year 2009. Master of Technology from National Institute of Technology, Raipur, India in the year of 2011. Now, working as an Assistant Professor at Department of Computer Science & Engineering in Institute of Engineering & Management, Kolkata. His areas of interests are Data Mining, Cloud Computing and Cryptography & Network Security.

**Suman Das:** B. Tech. from West Bengal University of Technology, India on Computer Science & Engineering in the year 2014. His areas of interests are Networking and Network Security.